\documentclass[12pt]{article}
\usepackage{amsmath}
\usepackage{amssymb}
\usepackage{array}
\usepackage{graphicx}
\usepackage{latexsym}
\usepackage{textcomp}
\usepackage{color}
\usepackage{soul}
\usepackage{blkarray}

\newtheorem{theorem}{Theorem}[section]

\newtheorem{corollary}{Corollary}[theorem]
\newtheorem{lemma}[theorem]{Lemma}

\newtheorem{definition}{Definition}[section]

\begin{document}

\title{Nonlinear learning and learning advantages in evolutionary games \thanks{This research is funded by the Australian Research Council Discovery Grants DP160101236 and DP150100618.}
}

\author{Maria Kleshnina \footnote{Centre for Applications in Natural Resource Mathematics (CARM), School of Mathematics and Physics, University of Queensland, St Lucia, Queensland 4072, Australia}, Jerzy A. Filar $^{\dagger}$, \\ 
Cecilia Gonzalez Tokman \footnote{School of Mathematics and Physics, University of Queensland, St Lucia, Queensland 4072, Australia}
}

\maketitle

\begin{abstract}
The idea of incompetence as a learning or adaptation function was introduced in the context of evolutionary games as a fixed parameter. However, live organisms usually perform different nonlinear adaptation functions such as a power law or exponential fitness growth. Here, we examine how the functional form of the learning process may affect the social competition between different behavioral types. Further, we extend our results for the evolutionary games where fluctuations in the environment affect the behavioral adaptation of competing species and demonstrate importance of  the starting level of incompetence for survival. Hence, we define a new concept of learning advantages that becomes crucial when environments are constantly changing and requiring rapid adaptation from species. This may lead to the evolutionarily weak phase when even evolutionary stable populations become vulnerable to invasions.

\end{abstract}


\section{Introduction}
\label{intro}

The question of what are the conditions for a particular type of species to survive frequently arises in ecology. This question is relevant to different environmental, social, genetic and other conditions. Evolutionary game theory, a branch of game theoretical and ecological sciences, aims to answer that question \cite{Apaloo2009,Hofbauer2003,Apaloo1995,Nowak2006,Smith1973}. However, habitats of species are dynamical environments. For example, the most obvious changes in the environment are daily or seasonal fluctuations. Moreover, animals might also migrate due to different factors. The concept of evolutionary games under incompetence was introduced to model such social problems of species \cite{Kleshnina2017}.

The concept of incompetence was first introduced from a classical game theory perspective \cite{Filar2012}. It aims to capture behavioral mistakes of players that are making their choice of strategies when interacting with others. Mathematically it means that the probability of executing the chosen strategy is not necessarily equal to one. The set of all such probabilities makes up the incompetence matrix, which, in turn, perturbs the fitness matrix of the game. Such perturbations under the static parameter of incompetence were examined in \cite{Kleshnina2017}. This paper aims to study evolutionary games under incompetence and time-dependent adaptation of species.

The degree of incompetence was captured by a parameter $\lambda$ that varies from $0$ to $1$ that reflects the level of ``competence'' of species. That is, the better the animals are in their strategy execution, the closer this parameter is to $1$. Now we define $\lambda(t)$ as a function of time and study its impact on game dynamics depending on the functional form of the adaptation. 

The adaptation processes of different live organisms have been studied widely in different fields \cite{Hoffmann2011,Losos2009,Sax2007}. Lenski et al. ran an experiment on the adaptation processes in Escherichia coli \cite{Lenski2009,Lenski1991,Lenski2000} and showed that after a population of bacteria was placed in a new environment, it evolved from low relative fitness to an adaptive peak or a plateau. Hence, the adaptation trajectory $\lambda(t)$ can be assigned to the entire population when the environment is static. However, in nature environments are changing due to different reasons such as seasonal fluctuations, climate changes, catastrophes or anthropogenic impacts. If unforeseen events are usually modeled as some form of a stochastic process, seasonal fluctuations were modeled via evolutionary game theory approach  \cite{Beninca2015}. In this paper we address two different adaptation trajectories and examine their influence on the evolutionary impact with the aim to find the adequate adaptation level to the fluctuating environment. Further, we introduce a new concept of learning advantages where the randomization of behavioral reactions might become crucial for surviving. Hence, we shall say that flexibility in reactions to the environmental changes is an important skill for animals living in fluctuating environments. 

\section{Mathematical model}
\label{mathmodel}

The idea of incompetence was introduced into classical replicator dynamics \cite{Kleshnina2017,Taylor1978} by considering a one population matrix game. In a classical sense, each player chooses an action (a pure strategy) and that choice results in a deterministic payoff: there is an underlying assumption that players are able to execute the actions that they have chosen. However, we assume the actions selected by players may not coincide with the executed actions and probabilities of actions execution form the incompetence matrix, $Q$, for the population \cite{Beck2013}. This stochastic matrix is made up from the set of all  probabilities that player 1 executes action $j$ given that he selects action $i$, and is given by:
\begin{equation}
\label{Qmatrix}
Q=\left(\begin{array}{cccc}
q_{11}& q_{21}& ...& q_{n1}\\
\vdots& \vdots& \ddots& \vdots\\
q_{1n}& q_{2n}& ...& q_{nn}
\end{array} \right).
\end{equation}

Then, we consider the expected reward matrix as a perturbed by incompetence fitness matrix in the form
\begin{equation}
\label{IncompetenceRM}
R^Q=Q R Q^T. 
\end{equation}

The way we construct a new incompetent fitness matrix $R^Q$ can be interpreted in the following way. At first, we shall consider pairwise interactions in a given population of animals that are immersed into new environmental conditions. These animals obtain a finite number, $n$, of available behavioral strategies. Hence, both interacting individuals are making their strategical choice which then leads to some payoff according to the fitness matrix $R$. However, we also assume that both individuals may be non-perfect in their strategy execution and, hence, can make mistakes with probabilities given by a matrix of incompetence, $Q$. Thus, both interacting individuals are prone to making behavioral mistakes which lead us to the expected payoff $r_{ij}^Q$, given that individuals are choosing strategies $i$ and $j$, defined as follows:

\begin{equation*}
\label{IncompetentReward}
r_{ij}^Q=\sum_{k=1}^{n} \sum_{h=1}^{n} q_{ik}q_{hj} r_{kh}, \;i,j=1,..,n,
\end{equation*}
where $q_{ij}$ denotes the $(i,j)^{\text{th}}$ entry of $Q$.

Furthermore, it is natural to assume that species are able to learn and adapt to the changing environment over time. Hence, we let the population to explore their environment via a learning process with an incompetence function $\lambda(\tau)$, that is we let them adapt from some starting level of incompetence to the perfect competence by the nonlinear adaptation law given by 
\begin{equation}
\label{incomp}
Q(\lambda(\tau))=(1-\lambda(\tau)) S+\lambda(\tau) I,\; \lambda(\tau)\in[0,1],
\end{equation}
where $S$ is the starting level of incompetence, $I$ is the identity matrix and $\tau$ is the learning time. 

Hence, in the evolutionary sense, behavioral mistakes lead to the perturbations in the fitness that the population obtains over time. This might be the case for the population migrating to the new unexplored environment or with the changing environment. In other words, when species have to adapt to new conditions. Hence, this process is immersed in the population dynamics: the fitness of $i$-th strategy is now given by
\begin{equation*}
\label{IncompFitness}
f_i(\lambda(\tau))=\sum_{j=1}^{n} r_{ij}(\lambda(\tau))x_j=e_i^T R(\lambda(\tau)) x,
\end{equation*}
where $R(\lambda(\tau))$ denotes an incompetent fitness matrix (\ref{IncompetenceRM}) with a time-varying incompetent matrix (\ref{incomp}), and the mean fitness of the entire population is defined as follows
\begin{equation}
\label{IncompMeanFitness}
	\phi(\lambda(\tau))=\sum_{i=1}^{n} x_if_i(\lambda(\tau))=x^T R(\lambda(\tau)) x.
\end{equation}

Based on these changes to the fitnesses we are now working with the incompetent replicator dynamics 
\begin{equation*}
	\dot{x}_i=x_i(f_i(\lambda(\tau))-\phi(\lambda(\tau))),\;i=1,...,n, 
\end{equation*}
or in a matrix form
\begin{equation}
\label{IncomReplicatorDynamics}
	\dot{x}_i=x_i((R(\lambda(\tau))x)_i-x^T R(\lambda(\tau)) x).
\end{equation}

That is, we explore the non-autonomous $n$-dimensional system. An important feature of such systems is that the time scale of replicator dynamics for $x(t)$ might not coincide with the time scale of adaptation for $\lambda(\tau)$. This means that individuals may explore their environment much faster or much slower than they reproduce, depending on the environment and particular species. We study this phenomenon in the Section \ref{sigmoid} when assuming a sigmoid learning process for the individuals. We also examine the periodic environmental fluctuations caused by the seasonal changes immersed in the evolutionary game via periodic forcing affecting the selection in Section \ref{periodic}. In Section \ref{learningadvantages} we introduce and explore a new concept of \emph{learning advantages} and study how that may affect adaptability of populations.


\section{Individual nonlinear learning}
\label{sigmoid}


We know that in the linear case, when $\lambda$ is treated as a fixed parameter, we observe bifurcations in the replicator dynamics that also depend on the starting level of incompetence $S$ \cite{Kleshnina2017}. In this section we will explore what is happening with the population dynamics when the process of learning is treated as a function of time. There are many options possible when we are choosing the form of the time-dependence. However, firstly we explore a sigmoid learning. This functional form is a reasonable choice for modeling individuals' learning process. For example, in bacterial adaptation to the new environment fitness of bacteria was slowly developing in the beginning with the following rapid growth and slowing down after some time \cite{Lenski1994}. This functional form also reflects the widely applied logistic growth of the population \cite{Gilpin1973}. Mathematically this function can be written as
\begin{equation*}
\lambda(\tau)=\frac{1}{1+e^{-a(\tau-b)}},
\end{equation*}
where parameters $a$ and $b$ depend on the initial conditions of the learning equation. We assume that learning time-scale $\tau$ is different from the reproduction time $t$ as species may reproduce with different pace from the one that they are exploring their environment. Let us assume that the learning time $\tau$ is scaled to the time of reproduction with some positive constant $\alpha$, that is, $\tau=\alpha t$. Then, we can rewrite the adaptation function as a function of time $t$

\begin{equation}
\label{lambda}
\lambda(t)=\frac{1}{1+e^{-a(\alpha t-b)}}.
\end{equation}

The adaptation process described in (\ref{incomp}) with fixed parameter $\lambda$ causes bifurcations in replicator dynamics and, hence, changes the outcome of the game. However, in the case with the incompetence parameter being a function of time we arrive into a new evolution process that depends on the parameters $a,b,\alpha$ from (\ref{lambda}). That is, the convergence to the expected outcome might be extended in time depending on the learning process. We will demonstrate this on the following example.

Consider a two-dimensional classical Hawk-Dove game with the fitness matrix:
$$\begin{blockarray}{ccc}
& Hawk & Dove \\
\begin{block}{c(cc)}
Hawk&-1&2\\
\\
Dove&0&1\\
\end{block}
\end{blockarray}$$

This game has an evolutionary stable strategy (ESS)  $\mathbf{\tilde{x}}=(\frac{1}{2},\frac{1}{2})$ \cite{Smith1973}. Let us introduce incompetence in this game with the starting level of incompetence as:
\begin{equation*}
S=\left(\begin{array}{ccc}
0.3 & & 0.7\\
 & & \\
0.6 & & 0.4
\end{array} \right).
\end{equation*}

Let us also introduce a learning process (\ref{incomp}) with the incompetence function from (\ref{lambda}). Dynamical behavior of the solution $\mathbf{x}(t)$ depending on parameters $a,b,\alpha$ is captured on Figure 1. Here we see that once being incompetent the population tends to behave differently from the fully competent case.

\begin{figure}[h!]
\begin{center}
\includegraphics[scale=0.35]{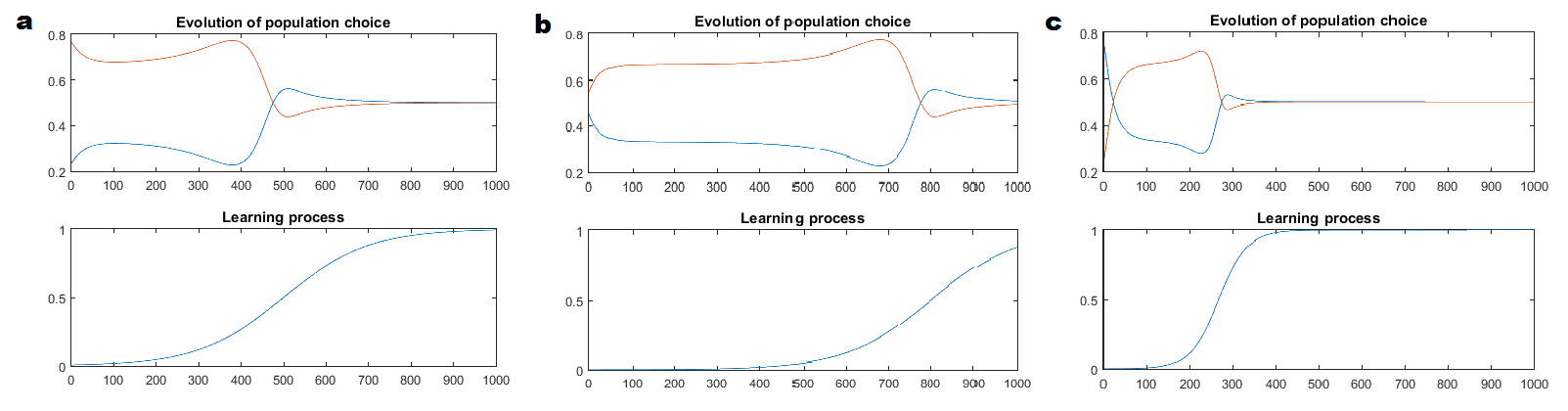}
\end{center}
\caption{Population dynamics depending on the configurations of the parameters $a,b,\alpha$ in $\lambda(t)$.}
\end{figure}

We observe that the impact of the learning process is higher when the population has just started to explore their environment. That is, bifurcations cause the dominance of the strategy that was unfavorable in the fully competent case and if the time scale of learning is large enough, this may lead to the situation where this strategy has an opportunity to outcompete another strategy. But according to (\ref{Qmatrix}) the execution of this strategy will still be perturbed and another strategy will reappear in the population as mistakes. Hence, this may become a tool for surviving. And even if the population has fully adapted (that is, $\lambda=1$), under the assumption that species may control their ability to execute different strategies, this may lead to a stable population. However, mathematically speaking, changes in dynamical behavior caused by the changes in the parameter, as time varies are called time-dependent bifurcations. In such a case, besides the bifurcation values of the parameter it is becoming important when parameter hits those values. In other words, we are interested in computing the critical time when bifurcations happen.

In our settings the time scale of learning,  $\alpha$, can be considered as a selection pressure or, even more, the measure of evolutionary success or simply a measure of fitness. By saying this, we mean that the faster species learn and adapt, the better their chance to reproduce and survive. Specifically, it is the case for the games with ESS when reaching an evolutionary stable state becomes a guarantee of survival. For example, speaking about bacteria, we say that {\itshape E.coli} live in almost static environment and would have a very small learning time-scaling constant $\alpha$ \cite{Lenski1994}. At the same time, marine bacteria might be required to adapt much faster due to the turbulence which causes changes in the environment \cite{Stocker2012}.

\subsection{Time-dependent bifurcations and critical time}

Let us first introduce some notation. Consider the replicator dynamics for the case with full competence, that is, the original evolutionary game. We shall define such a game as $\Gamma_1$, the fully competent game\footnote{In (\ref{fullcompetent})-(\ref{lambdavarying}), and elsewhere, we suppress the argument, $t$, of differentiation in line with common practice in the field}:

\begin{equation}
\label{fullcompetent}
\Gamma_1 : \; \dot{\mathbf{x}}=\mathbf{g}(\mathbf{x},1)=\mathbf{g}(\mathbf{x}), \; \mathbf{x}(0)=\mathbf{x}_0.
\end{equation}

Further, we define a game  with the fixed incompetence parameter, $\lambda$, as a $\lambda$-fixed incompetent game $\Gamma_\lambda$: 

\begin{equation}
\label{fixedlambda}
\Gamma_\lambda : \; \dot{\mathbf{x}}=\mathbf{g}(\mathbf{x},\lambda), \; \mathbf{x}(0)=\mathbf{x}_0.
\end{equation}

Next, we shall call a $\lambda(t)$-varying game $\Gamma_{\lambda(t)}$ with $\lambda$ being time-dependent as follows.

\begin{equation}
\label{lambdavarying}
\Gamma_{\lambda(t)} : \; \dot{\mathbf{x}}=\mathbf{g}(\mathbf{x},\lambda(t)), \; \mathbf{x}(0)=\mathbf{x}_0.
\end{equation}

For this section we assume the form of the function $\lambda(t)$ from (\ref{lambda}). Let us define \emph{a maximal critical value of} $\lambda$ as in \cite{Kleshnina2017}.

\begin{definition}
A maximal critical value of the incompetence parameter, $\lambda^u \in [0,1]$, is the maximal bifurcation parameter value, $\lambda^c$, for the fixed point of incompetent games $\Gamma_\lambda$.
\end{definition}

Namely, for any $\lambda>\lambda^u$ the dynamics of $\Gamma_\lambda$ remain qualitatively similar to the $\Gamma_1$ game. Let us assume that the system $\Gamma_\lambda$ has a stable equilibrium $\tilde{\mathbf{x}}(\lambda)$ for any $\lambda\in [\lambda^u,1]$ with a fixed point $\tilde{\mathbf{x}}(\lambda)$ \cite{Kleshnina2017}. From (\ref{lambda}) we can identify the time $t^u$ such that $\lambda(t^u)=\lambda^u$. Next, let us define the asymptotic stability of the fixed point $\mathbf{\tilde{x}}(\lambda)$.



\begin{definition}
Assume $\lambda\in [\lambda^u,1]$. An equilibrium $\mathbf{\tilde{x}}(\lambda)$ of $\Gamma_\lambda$ is called asymptotically stable if for any $\epsilon >0$ there exists $\delta(\epsilon)>0, t^\epsilon \geq t^u$ such that the solution $x(t)$ of $\Gamma_{\lambda(t)}$ with $x(0)=x_0$ satisfies
$$||x(t)-\mathbf{\tilde{x}}(\lambda)||<\epsilon,\;\forall t\in(t^\epsilon,\infty),$$
when $||x_0-\mathbf{\tilde{x}}(\lambda)||<\delta(\epsilon)$.
\end{definition}


Let us also recall Theorem 1 from \cite{Kleshnina2017} that the existence of an ESS in the original game implies the existence of an ESS in the incompetent game after some $\lambda^u$.

\begin{theorem}
\label{ESStheorem}
Let $\tilde{x}$ be an ESS of the fully competent game $\Gamma_1$. Let $\lambda^u\in [0,1]$ be the maximal critical value of $\Gamma_\lambda$. Suppose that $||Q(\lambda)-I||\leq \delta(\lambda^u)$, when $\lambda\in (\lambda^u,1]$ and $\delta(\lambda^u)$ is sufficiently small, then the incompetent game $\Gamma_\lambda$ possesses an ESS $\tilde{x}(\lambda)$ and $$\lim_{\lambda\rightarrow 1^{-}} \tilde{x}(\lambda) = \tilde{x}.$$
\end{theorem}

In addition, we require the next lemma based on the proof of Theorem 1 in \cite{Zeeman1980}.

\begin{lemma}
\label{LyapunovZeeman}
Let $\tilde{x}$ be an ESS of the fully competent game $\Gamma_1$. Then the function $V(\mathbf{x}) = \prod_{i=1}^n x_i ^{\tilde{x}_i}$ is a strict local Lyapunov function for the replicator dynamics with the derivative along the trajectories being $\dot{V}(\mathbf{x}) = V(\mathbf{x}) \left( \mathbf{\tilde{x}}R\mathbf{x} - \mathbf{x}R\mathbf{x} \right)>0, \forall \mathbf{x}\neq \mathbf{\tilde{x}}$, x in a neighborhood $\mathcal{N}$ of $\mathbf{\tilde{x}}$.
\end{lemma}

Then the following result can now be established.

\begin{theorem}
\label{ESSindividual}
Let $\tilde{\mathbf{x}}:=\tilde{\mathbf{x}}(1)$ be an ESS for $\Gamma_1$. Now consider the $\lambda(t)$-varying game $\Gamma_{\lambda(t)}$, where $\lambda(t)$ is the sigmoid adaptation function (\ref{lambda}). Let $\lambda^u>\lambda^c$ for all bifurcation points $\lambda^c$ of the $\lambda$-fixed games $\Gamma_\lambda$ and $t^u>0$ be the time when $\lambda(t^u)=\lambda^u$. Then there exists $t>t^u$ such that $\tilde{\mathbf{x}}$ is an ESS for $\Gamma_{\lambda(t)}$ on the interval $(t,\infty)$.
\end{theorem}

{\bf Proof}: An ESS is an asymptotically stable fixed point of the replicator dynamics \cite{Bomze1986}. 
From (\ref{incomp}) we have

\begin{equation*}
\label{QS}
Q(\lambda(t)) = (1-\lambda(t))S+\lambda(t)I,
\end{equation*}

which from (\ref{IncompetenceRM}) yields

\begin{align}
\label{RS}
& R(\lambda(t)) = [(1-\lambda(t))S+\lambda(t)I]R[(1-\lambda(t))S+\lambda(t)I]^T \\ \nonumber
& = \lambda(t)^2 R + \lambda(t)(1-\lambda(t))[SR+RS^T]+(1-\lambda(t))^2 SRS^T.
\end{align}

Substituting (\ref{RS}) into (\ref{IncomReplicatorDynamics}) we obtain

\begin{align}
\label{RDS}
&\mathbf{g}(\mathbf{x},t)= \lambda(t)^2 \mathbf{g}(\mathbf{x}) +\lambda(t)(1-\lambda(t)) \mathbf{g}^{SS^T}(\mathbf{x},t) + (1-\lambda(t))^2\mathbf{g}^{S} (\mathbf{x},t) ,
\end{align}
with $\mathbf{g}^{SS^T} (\mathbf{x},t) = X \left( [SR+RS^T]\mathbf{x} - \mathbf{x}^T [SR+RS^T] \mathbf{x} \mathbf{1} \right)$ and \\ $\mathbf{g}^{S} (\mathbf{x},t) = X \left( SRS^T\mathbf{x} - \mathbf{x}^TSRS^T\mathbf{x} \mathbf{1} \right)$, where $X$ is a diagonal matrix with $\mathbf{\tilde{x}}$ on the diagonal and $\mathbf{1}$ is a vector of ones.

Consider the function $V(\mathbf{x}) = \prod_{i=1}^n x_i ^{\tilde{x}_i}$ as a candidate for the strict local Lyapunov function of the equation (\ref{RDS}) according to Lemma 1, then

\begin{align*}
& \dot{V}(\mathbf{x},t) = \lambda(t)^2 V(\mathbf{x}) \left( \mathbf{\tilde{x}}R\mathbf{x} - \mathbf{x}R\mathbf{x} \right) + (1-\lambda(t))^2 ( V(\mathbf{x}) \left( \mathbf{\tilde{x}}SRS^T\mathbf{x} - \mathbf{x}SRS^T\mathbf{x} \right) + \\
& \lambda(t)(1-\lambda(t)) V(\mathbf{x}) \left( \mathbf{\tilde{x}}[SR+RS^T]\mathbf{x} - \mathbf{x}[SR+RS^T]\mathbf{x} \right) ) = \lambda(t)^2 \dot{V}(\mathbf{x})+h(\mathbf{x},\lambda(t)).
\end{align*}

Then, as $\lambda(t)\rightarrow 1$, $h(\mathbf{x},\lambda(t)) \rightarrow 0$ and for a every $\epsilon >0$ there exists $t^\epsilon > t^u$ such that, 

$$||\dot{V}(\mathbf{x},t) - \dot{V}(\mathbf{x})||<\epsilon,$$
provided that $\mathbf{x}$ is in the attraction region $\mathcal{N}$ of $\mathbf{\tilde{x}}$. In order to show that $\tilde{\mathbf{x}}$ is an ESS of $\Gamma_{\lambda(t)}$ on $(t,\infty)$ it is sufficient to show that for every $x\in \mathcal{N}$ and $t>t^\epsilon$, either  $\dot V(x,t) >0$ or $|x-\tilde{x}|<\delta(\epsilon)$ for some $\delta(\epsilon)$ such that
$\lim_{\epsilon \rightarrow 0} \delta(\epsilon) = 0$. Recall that $\dot{V}(\mathbf{x}) = V(\mathbf{x})(\mathbf{\tilde{x}}R\mathbf{x} - \mathbf{x}R\mathbf{x})$. As $V(\mathbf{x})$ is a strict local Lyapunov function, then $V(\mathbf{x})>0$. Next, let us consider the function $\mathbf{\tilde{x}}R\mathbf{x} - \mathbf{x}R\mathbf{x}$, which is equal to $0$ only at $\mathbf{\tilde{x}}$ and is positive for any other $\mathbf{x}$ by the definition of ESS. Then, for every $\epsilon>0$ sufficiently small there exists $\delta(\epsilon)$ such that the preimage of $[0,\epsilon)$ under $\dot{V}(\mathbf{x})$ satisfies $\dot{V}^{-1}([0,\epsilon)) \cap \mathcal{N} \subset \mathcal{B}_{\delta(\epsilon)}(\tilde{x})$ and $\lim_{\epsilon \to 0} \delta(\epsilon)=0$. This completes the proof.

\hfill $\square$

Let us now define {\itshape the critical time} of the learning adaptation process as the first time when incompetence function, $\lambda(t)$, attains the maximal critical value, $\lambda^u$, of Theorems \ref{ESStheorem}-\ref{ESSindividual}. We shall say that the species is in an {\itshape evolutionarily weak phase} prior to $t^u$ because it is susceptible to invasions that may prevent it from reaching the ESS, $\tilde{\mathbf{x}}$.  

\begin{corollary}
\label{crittime}

Let $\lambda^u$ be the maximal bifurcation value of the incompetence parameter for $\mathbf{\tilde{x}}$, then the critical time is given by 

\begin{equation}
\label{tcrit}
t^u = \frac{1}{\alpha} \left( b-\frac{1}{a} \ln\left(\frac{1-\lambda^u}{\lambda^u} \right) \right).
\end{equation}


\end{corollary}

{\bf Proof:}
Proof of this corollary is following directly from the functional form of $\lambda(t)$ given by (\ref{sigmoid}). Namely, we solve for $t^u$ the equation

$$\lambda^u=\frac{1}{1+e^{-a(\alpha t^u-b)}}.$$ 

\hfill $\square$







Individual learning describes the adaptation process of the entire population to the new environment. This functional form of $\lambda(t)$ can be applied to the analysis when species are immersed in a new environment due to the migration process or some abrupt change. By analyzing parameters of $\lambda(t)$ we can say how long will it take for species to be fully recovered in behavioral sense and act as in the environment they are familiar with. However, while the functional form of the adaptation trajectory captures the pace and steepness of the learning process, the starting level of incompetence measures how dramatic the changes in the environment are. That is, the more different is  the new habitat from the previous one, the longer it may take for organisms to fully recover. 

Individual learning covered in this section describes adaptation to once off change in species' habitat. However, natural environments are prone to some stochasticity or seasonal fluctuations \cite{Beninca2015,Fuhrman2015,Gilbert2012,Suttle2007}. In order to model this, we may use periodic functions for the learning process of organisms. 


\section{Environmental fluctuations: periodic forcing}
\label{periodic}


Changing environment was modeled with help of evolutionary game theory as a stochastic process \cite{Foster1990}. However, stochasticity is not the only factor characterizing environmental changes. Seasonal fluctuations in temperature, humidity, food availability all tend to cause periodic environmental changes. Depending on the regional factors, these fluctuations can be almost unnoticeable or extreme. Hence, different habitats may require different adaptation skills from living organisms \cite{Fuhrman2015,Gilbert2012}.

Mathematically, that means that the incompetence parameter need not rise from $0$ towards $1$ just once when the environment has changed permanently. Instead, in synchrony with the aim of capturing seasonal fluctuations, the adaptation function is now endowed with a shape of a periodic function. Mathematically, we start with the functional form.

\begin{equation}
\label{periodiclambda}
\lambda(t) = \frac{1}{2} \sin(\alpha t) + \frac{1}{2}, \; t\in(0,\infty),
\end{equation}
where $\alpha$ is interpreted as a frequency of fluctuations in the environmental conditions. For instance, small $\alpha$ reflects longer period after which the learning cycle repeats itself. For example, see Figure 2.

\begin{figure}[h!]
\begin{center}
\includegraphics[scale=0.55]{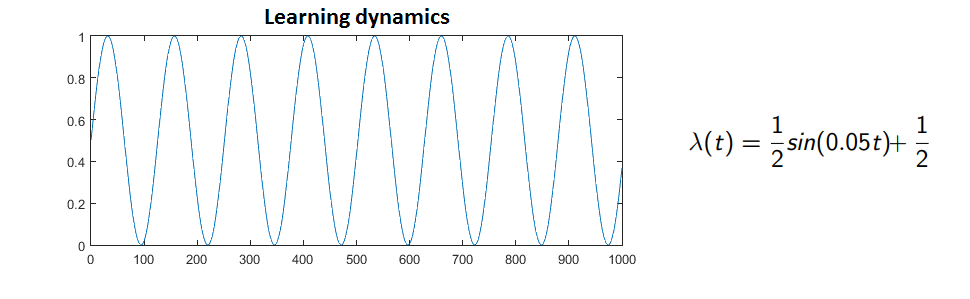}
\end{center}
\caption{Learning function $\lambda(t)$ for $\alpha=0.05$ and period of approximately 125 days.}
\end{figure}

We start with the sine function for now, however, this functional form could have a more complicated shape reflecting different degrees of changes between seasons.

Let us now demonstrate how periodic forcing changes the selection process on the classical example of the Hawk-Dove-Retaliator game. This game was widely studied (e.g. see \cite{Bomze1983}) and it was shown in \cite{Kleshnina2017} that incompetence causes bifurcations of the game's fixed points. The fitness matrix is constructed as:
$$\begin{blockarray}{cccc}
& Hawk & Dove & Retaliator\\
\begin{block}{c(ccc)}
Hawk&-1&2&-1\\
\\
Dove&0&1&1\\
\\
Retaliator&-1&1&1\\
\end{block}
\end{blockarray}.$$

\begin{figure}[h!]
\begin{center}
\includegraphics[scale=0.55]{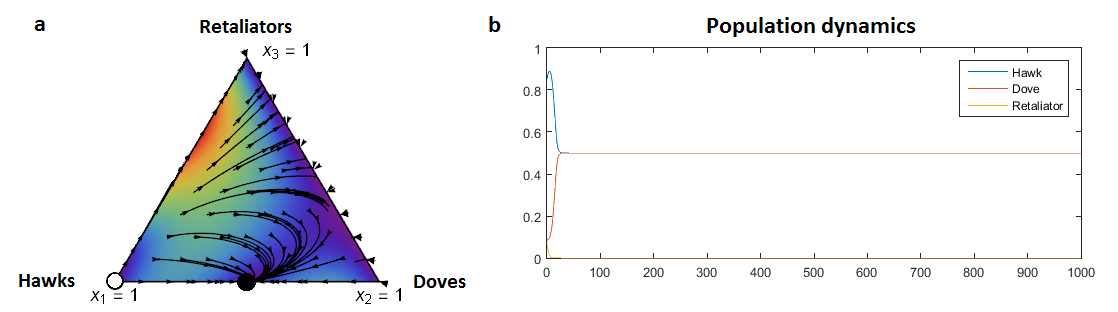}
\end{center}
\caption{Game flow (a) and population dynamics (b) for Hawk-Dove-Retaliator game.}
\end{figure}

\begin{figure}[h!]
\begin{center}
\includegraphics[scale=0.55]{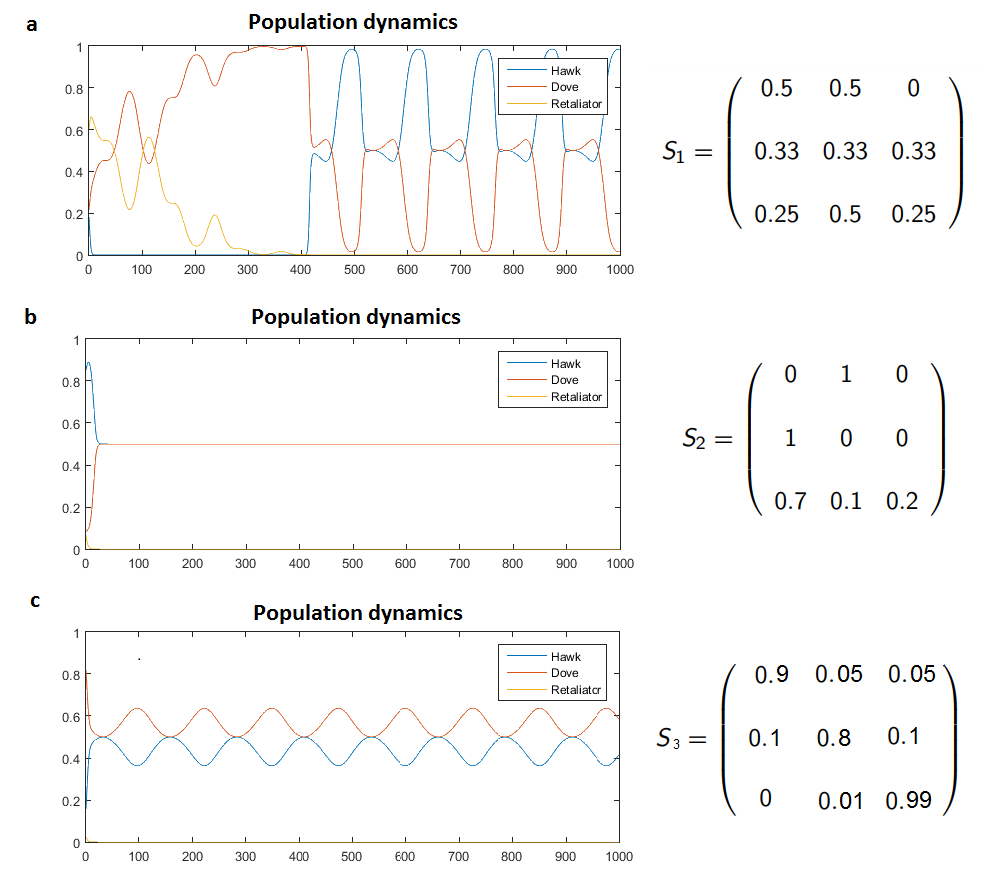}
\end{center}
\caption{Population dynamics for the Hawk-Dove-Retaliator game for three different starting levels of incompetence corresponding to one periodic adaptation trajectory $\lambda(t)$: (a) the starting level of incompetence $S_1$; (b) the starting level of incompetence $S_2$; (c) the starting level of incompetence $S_3$.}
\end{figure}

Population dynamics of the original fully competent game are captured in Figure 3. We know that the original Hawk-Dove-Retaliator game possesses an ESS which is a fair mixture of Hawks and Doves. In a case with the fixed parameter $\lambda$ we obtain different game flows following from the values of that parameter. It depends on the structure of the starting level of incompetence matrix, $S$. 

Dynamics of the incompetent game that uses (\ref{sigmoid}) are depicted in Figure 4. Figure 4 (a)-(b) demonstrates two population trajectories that correspond to two different $S$ matrices $S_1$ and $S_2$, respectively. The first matrix reflects greater mixing of the behaviors for all three strategies, whereas the second matrix reflects mixing only for Retaliators, while Hawks and Doves just interchange their behavior. The first matrix, $S_1$, induces periodic behavior of the solution whereas the second matrix, $S_2$, does not cause any change in the population dynamics, compare to Figure 3 (b). 

\newpage

Note that, in view of (\ref{periodiclambda}) and (\ref{incomp}) 

$$Q(\lambda(t)) = S,\; \text{when } t=\frac{2 \pi k}{\alpha},\; k=0,1,2,\ldots .$$

Thus the learning time scale period is $\frac{2\pi}{\alpha}$. Hence, in the case with periodic forcing the impact of $S$ is becoming even more significant. The starting level of incompetence here may represent variations in the environmental changes. 




Obviously, we see that existence of critical values of the incompetence parameter is influenced by both the starting level of incompetence, $S$, and a fitness matrix, $R$. In addition, for the sufficiently small perturbations, that is, when $S$ is sufficiently close to the identity matrix, we may observe periodic behavior in the dynamics of games with ESS. Below, we use a result from \cite{Hale1991} which states that a hyperbolic equilibrium solution of an autonomous differential equation persists as a periodic cycle under small periodic perturbations of the parameter. In a strict sense this statement is formulated as follows, with subscripts $x$ and $t$ denoting partial derivatives with respect to these arguments.


\begin{theorem}\cite{Hale1991}
\label{HaleTheorem}
Let $\dot{x}=G(x)$ be an autonomous differential equation, with $G(\tilde{x})=0,\;G'(\tilde{x})\neq 0,$ and $x_1(t)=\tilde{x}$ the resulting equilibrium solution. Consider $\dot{x}=G(x,t)$ where $G(x,t)$ is periodic in $t$ with period $T$. If for all $(x,t)$, $|G(x,t)-G(x)|$, $|G_x(x,t)-G'(x)|,$ and $|G_t(x,t)|$ are sufficiently small, then there is a periodic solution $x_2(t)$ of the time-dependent equation that stays arbitrarily close to the solution of the autonomous equation.
\end{theorem}







Now, using Theorem \ref{HaleTheorem} and the fact that $\mathbf{g}(\mathbf{x},\lambda(t))$ in (\ref{lambda}) is periodic the following result can be formulated.

\begin{theorem}
\label{TheoremPeriodic}
Let $\tilde{\mathbf{x}}$ be an ESS of $\Gamma_1$ and $x_1(t)$ be a resulting solution for some $\mathbf{x}_1(0)=\mathbf{x}_0$. For $\lambda(t)$ being a periodic function of period $T=\frac{2\pi}{\alpha}$ from (\ref{periodiclambda}) and a sufficiently small $\delta$ such that $||S-I||<\delta$ there exists a periodic solution $x_2(t)$ with $\mathbf{x}_2(0)=\mathbf{x}_0$ of $\Gamma_{\lambda(t)}$ that stays arbitrarily close to $x_1(t)$. 
\end{theorem}

{\bf Proof:}
In order to apply Theorem \ref{HaleTheorem} we need to estimate $|\mathbf{g}(\mathbf{x},t)-\mathbf{g}(\mathbf{x})|,$ $|\mathbf{g}_x(\mathbf{x},t)-\mathbf{g}'(\mathbf{x})|,$ and $|\mathbf{g}_t(\mathbf{x},t)|$, and show that they are sufficiently small. The first distance $|\mathbf{g}(\mathbf{x},t)-\mathbf{g}(\mathbf{x})|$ is sufficiently small due to the Lipschitz continuity of the replicator dynamics (see p. 141 \cite{Weibull1997}).

Next, we can rewrite the requirement $||S-I||<\delta$ as $S=I+\mathcal{E}(\delta)$, where $\mathcal{E}(\delta)$ is a matrix with entries $\epsilon_{ij}(\delta)$ that are sufficiently small. 
For the simplicity of notation we omit the dependence on $\delta$ and simply use $\mathcal{E}$. From (\ref{incomp}) we have

\begin{equation*}
\label{Qepsilon}
Q(\lambda(t)) = (1-\lambda(t))(I+\mathcal{E})+\lambda(t)I = I + (1-\lambda(t))\mathcal{E},
\end{equation*}

which from (\ref{IncompetenceRM}) yields

\begin{align}
\label{Repsilon}
R(\lambda(t)) = [I + (1-\lambda(t))\mathcal{E}]R[I + (1-\lambda(t))\mathcal{E}]^T \\ \nonumber
= R + (1-\lambda(t))[\mathcal{E}R+R\mathcal{E}^T]+(1-\lambda(t))^2 \mathcal{E}R\mathcal{E}^T.
\end{align}

Substituting (\ref{Repsilon}) into (\ref{IncomReplicatorDynamics}) we obtain

\begin{align*}
&\mathbf{g}(\mathbf{x},t)= \mathbf{g}(\mathbf{x}) + \mathbf{g}^\epsilon(\mathbf{x}, \lambda(t)),
\end{align*}
where $\mathbf{g}^\epsilon(\mathbf{x}, \lambda(t)) = (1-\lambda(t))^2 X \left( \mathcal{E}R\mathcal{E}^T\mathbf{x} - \mathbf{x}^T \mathcal{E}R\mathcal{E}^T \mathbf{x} \mathbf{1} \right) + \\ (1-\lambda(t)) X \left( [\mathcal{E}R+R\mathcal{E}^T]\mathbf{x} - \mathbf{x}^T[\mathcal{E}R+R\mathcal{E}^T]\mathbf{x} \mathbf{1} \right)$. Here, $X$ is a diagonal matrix with $\mathbf{\tilde{x}}$ on the diagonal.
Hence, $|\mathbf{g}_x(\mathbf{x},t)-\mathbf{g}'(\mathbf{x})| = |\mathbf{g}^\epsilon_x(\mathbf{x}, \lambda(t))|$ and for any $\mathcal{E}$ with $|\epsilon_{ij}| << 1$ there exists $\sigma$ such that $|\mathbf{g}_x(\mathbf{x},t)-\mathbf{g}'(\mathbf{x})| < \sigma$.

Next, note that 

$$\mathbf{g}_t(\mathbf{x},t) = \dot{\lambda} (t)\frac{\partial \mathbf{g}^{\epsilon}(\mathbf{x},t)}{\partial \lambda(t)} = \frac{1}{2} \alpha cos(\alpha t)  \frac{\partial \mathbf{g}^{\epsilon}(\mathbf{x},t)}{\partial \lambda(t)},$$

where 

\begin{align*}
\frac{\partial \mathbf{g}^{\epsilon}(\mathbf{x},\lambda(t))}{\partial \lambda} =  -( 2(1-\lambda(t)) X \left( \mathcal{E}R\mathcal{E}^T\mathbf{x} - \mathbf{x}^T \mathcal{E}R\mathcal{E}^T \mathbf{x}\mathbf{1} \right) - \\
X \left( [\mathcal{E}R+R\mathcal{E}^T]\mathbf{x} - \mathbf{x}^T[\mathcal{E}R+R\mathcal{E}^T]\mathbf{x}\mathbf{1} \right).
\end{align*}

Hence, the derivative $\mathbf{g}_x(\mathbf{x},t)$ of the periodic replicator dynamics is also sufficiently small. 


\hfill $\square$

We have demonstrated that periodic environmental fluctuations may lead to the periodicity in the population dynamics (see Figure 4 (c) for the demonstration of Theorem \ref{TheoremPeriodic}). This idea was captured by the well-known Lotka-Volterra equations and many other ecological models \cite{Bazykin1998}. 

We explain the periodicity in species behavior from the game theoretical point of view when changing environment requires different levels of adaptation and flexibility from species needing to adapt. Consequently, the ability to react to the fluctuations becomes crucial. 

In the view of above, the concept of learning advantages is arising in the population dynamics analysis. We shall say that the ability to randomize behavior as a reaction to environmental changes may become crucial for species survival. In other words, it is not that important \emph{how} species adapt, but the ability to be \emph{flexible} is becoming crucial. The requirement to be flexible seems to be natural when we model environments that are prone to regular changes.

\section{Learning advantages}
\label{learningadvantages}

While it may appear counterintuitive at first, we shall say that strategies that are prone to mistakes have a {\itshape learning advantage}. By saying this, we mean that, via mistakes, species may obtain benefits of randomizing their behavioral patterns. This may enhance their chance to survive invasions of mutants and some strategies could be adopted faster than others. 


  In other words, we shall say that a strategy $i$ obtains a potential learning advantage if $S_i$ the $i$-th row of $S$ differs from $e_i$, where $e_i$ is the $i$-th member of a unitary basis, and the species of type $i$ are able to randomize their reaction to environmental changes according to $S_i$. We have chosen the word advantage to describe this difference from other strategies as species of $i$-th type are capable of executing strategies differing from those of their own type. In some games, this means that they are able to execute strategies \emph{required by the environmental conditions}. This ability may be crucial when we consider changing environments where the ability to adapt becomes particularly important. 

  Hence the degree of incompetence may constitute either an evolutionary disadvantage or an advantage. However, this depends strongly on the interplay among learning dynamics, learning advantages and fitness advantages. Let us demonstrate this by considering the particular form of the starting level of incompetence when no learning advantages are assumed for any strategy. 

\begin{definition}
A \emph{uniform incompetent game} is a game with the uniform matrix as a starting level of incompetence, that is, $s_{ij}=\frac{1}{n},\forall i,j=1,\ldots,n$. 
\end{definition}

Next, we shall recall an equivalence result on the positive affine transformation of the fitness matrix.

\begin{lemma}\cite{Bomze1986}
\label{transformation}
If $\hat{R}=k R+C_c$, where $k>0$ and $C_c$ is some column-constant matrix, then $\Gamma$ and $\hat{\Gamma}$ have identical ESS, fixed points and their stability properties.
\end{lemma}

We can then simplify our incompetent game in the following manner.

\begin{lemma}
If the starting level of incompetence, $S$, is a uniform matrix, that is, $s_{ij} = \frac{1}{n},\forall i,j=1,\ldots,n$, then the incompetent game $\Gamma_\lambda$ is equivalent to the simplified uniform incompetent game with the fitness matrix $\hat{R}(\lambda)=R+\frac{1-\lambda}{\lambda n} RJ$, where $J$ is a matrix of ones and $R$ is the fitness matrix of the original game.
\end{lemma}

{\bf Proof:}

Due to the fact that $S$ is a uniform matrix we can rewrite the matrix of incompetence (\ref{incomp}) as:

$$Q(\lambda)=\frac{1-\lambda}{n} J + \lambda I = Q^T(\lambda).$$

Then, the incompetent fitness matrix from (\ref{IncompetenceRM}) can be rewritten as

$$R(\lambda)=\left( \frac{1-\lambda}{n} \right) ^2 JRJ + \frac{\lambda(1-\lambda)}{n} \left( RJ + JR \right) + \lambda^2 R = \lambda^2 R + C_r + C_c,$$

where $JR$ is a column-constant matrix with the constant element of the $j$-th column being $r_{* j} = \sum_i r_{ij}$. Similarly, $RJ$ is a row-constant matrix with the constant element of the $i$-th row being $r_{i *} = \sum_j r_{ij}$. Finally, $JRJ$ is a constant matrix with all elements being $\sum_j \sum_i r_{ij}$. In the above, we have set $C_c:= \left( \frac{1-\lambda}{n} \right) ^2 JRJ + \frac{\lambda(1-\lambda)}{n} JR$, a row-constant matrix and $C_r:=\frac{\lambda(1-\lambda)}{n} RJ $ a column-constant matrix. Then the fitness matrix can be written as

\begin{equation}
\label{Rhat}
R(\lambda)=\lambda^2 (R + \frac{1}{\lambda^2}C_r + \frac{1}{\lambda^2}C_c). 
\end{equation}



Applying Lemma \ref{transformation} to (\ref{Rhat}) we see that it is sufficient to consider only the equilibria and fixed points of

\begin{equation}
\label{uniformR}
\hat{R}(\lambda)=R+\frac{1-\lambda}{\lambda n} RJ.
\end{equation}

\hfill $\square$

The equivalent fitness matrix (\ref{uniformR}) has a very specific form which provides a following result.

\begin{lemma}
\label{determinant}
If $\hat{R}(\lambda)=R+\frac{1-\lambda}{\lambda n} RJ$, then $\det(\hat{R}(\lambda))=\frac{1}{\lambda}\det(R)$.
\end{lemma}

{\bf Proof:}

(i) Let us first consider the matrix $\hat{R}=R+RJ$, hence, $\hat{r}_{ij}=r_{ij}+r_{i.}$. According to \cite{Filar1984} the determinant of such matrix can be written in the following form:

\begin{equation}
\label{detRhat}
\det(\hat{R})=\det(R)+\det(\mathbf{\tilde{c}},\mathbf{c}_2-\mathbf{c}_1,\ldots,\mathbf{c}_n-\mathbf{c}_1),
\end{equation}

where $\mathbf{c}_j$ are columns of $R$ and $\mathbf{\tilde{c}}$ is a vector of row-sums of $R$. Note that the second term of the determinant of $\hat{R}$ is a sum of $n$ determinants:

\begin{equation*}
\det(\mathbf{\tilde{c}},\mathbf{c}_2-\mathbf{c}_1,\ldots,\mathbf{c}_n-\mathbf{c}_1) = \sum_{j=1}^n \det(\mathbf{c}_j,\mathbf{c}_2-\mathbf{c}_1,\ldots,\mathbf{c}_n-\mathbf{c}_1).
\end{equation*}

Consider one of the determinants, in the above summation, for some $j$:

\begin{align*}
\det(\mathbf{c}_j,\mathbf{c}_2-\mathbf{c}_1,...,\mathbf{c}_j-\mathbf{c}_1,...,\mathbf{c}_n-\mathbf{c}_1) = \det(\mathbf{c}_j,\mathbf{c}_2-\mathbf{c}_1,...,-\mathbf{c}_1,...,\mathbf{c}_n-\mathbf{c}_1) \\ \nonumber
= -\det(-\mathbf{c}_1,\mathbf{c}_2-\mathbf{c}_1,...,\mathbf{c}_j,...,\mathbf{c}_n-\mathbf{c}_1) = \det(\mathbf{c}_1,\mathbf{c}_2,...,\mathbf{c}_j,...,\mathbf{c}_n) = \det(R).
\end{align*}

Hence, from (\ref{detRhat}) $\det(\hat{R})=(n+1)\det(R)$. 

(ii) Now, let us consider the matrix $\hat{R}(\lambda)=R+\frac{\lambda(1-\lambda)}{n} RJ$. Due to the fact that the second term of the determinant of $\hat{R}(\lambda)$ can be written as \\ $\det(\frac{1-\lambda}{\lambda n}\mathbf{\tilde{c}},\mathbf{c}_2-\mathbf{c}_1,\ldots,\mathbf{c}_n-\mathbf{c}_1)$, we obtain the required result.

\hfill $\square$

We know from \cite{Kleshnina2017} that there are a very special balanced bifurcations of the fixed points occurring for some values of $\lambda$. Let us first recall the definition of the balanced bifurcation point and the result for the interior fixed point:

\begin{definition}
We shall say that $\lambda^c$ is {\em a balanced bifurcation parameter value} of the fixed point $\tilde{\mathbf{x}}$ when it is a bifurcation of $\Gamma_\lambda$ and the mean fitness of the population $\phi(\tilde{\mathbf{x}},\lambda^c)=0$.
\end{definition}

\begin{lemma}\cite{Kleshnina2017}
If $\mathbf{\tilde{x}}$ is an interior fixed point of $\Gamma_1$, that is, $\tilde{x}_i>0,\forall i$, then every balanced bifurcation parameter value, $\lambda^c$, is also a singular point of $\tilde{R}(\lambda)$, that is $\det[\tilde{R}(\lambda)]=0$.
\end{lemma}

Hence, Lemmas 3-4 imply that for the case of the uniform incompetence we do not obtain any balanced bifurcation points except, perhaps, $\lambda=0$. However, we may extend our conclusions and formulate the following result:

\begin{lemma}
Let $\tilde{\mathbf{x}}$ be an interior ESS of $R$. If the starting level of incompetence, $S$, is a uniform matrix, that is, $s_{ij} = \frac{1}{n},\forall i,j=1,\ldots,n$ and $R$ is a row-sum-constant matrix, then $\tilde{\mathbf{x}}$ is an interior ESS for the incompetent game $\hat{R}(\lambda)$, for any $\lambda\in[0,1]$.
\end{lemma}

{\bf Proof:}

Since $\tilde{\mathbf{x}}$ is an interior ESS, by Proposition 12 in \cite{Bomze1986} it is the unique solution of the equation

$$R\tilde{\mathbf{x}}=(\tilde{\mathbf{x}}R\tilde{\mathbf{x}}) \mathbf{1},$$

where $\mathbf{1}$ is a vector of ones. Therefore, $R^{-1}$ exists and equals $\frac{1}{\det(R)} [R_{ij}]^T$, where $R_{ij}$'s are cofactors of $R$ and we have

\begin{equation}
\label{xk}
\tilde{x}_k=\frac{\det(R)}{\sum_{j=1}^n \sum_{i=1}^n R_{ij} } \times \frac{\sum_{j=1}^n R_{jk}}{\det(R)} = \frac{\sum_{j=1}^n R_{jk}}{\sum_{j=1}^n \sum_{i=1}^n R_{ij} }.
\end{equation}

From Lemma 1 we can consider the simplified uniform fitness matrix $\hat{R}(\lambda)=R+\frac{1-\lambda}{\lambda n} JR$. From Lemma 2.1 of \cite{Filar1984} we know that

$$\sum_{j=1}^n \sum_{i=1}^n R_{ij} = \sum_{j=1}^n \sum_{i=1}^n \hat{R}(\lambda)_{ij}.$$

Suppose now that $\tilde{\mathbf{x}}$ is not an interior ESS for the game with $\hat{R}(\lambda)$. Then,

$$\hat{R}(\lambda)\tilde{\mathbf{x}} \neq (\tilde{\mathbf{x}}\hat{R}(\lambda)\tilde{\mathbf{x}}) \mathbf{1},$$

and, hence, 

\begin{equation}
\label{xk2}
\tilde{x}_k \neq \tilde{x}_k(\lambda) = \frac{\sum_{j=1}^n \hat{R}(\lambda)_{jk}}{\sum_{j=1}^n \sum_{i=1}^n \hat{R}(\lambda)_{ij} }.
\end{equation}

However, according to the proof of Lemma 3.3 in \cite{Filar1984} we can rewrite the right-hand side of (\ref{xk2}) as

\begin{align*}
\frac{\sum_{j=1}^n \hat{R}(\lambda)_{jk}}{\sum_{j=1}^n \sum_{i=1}^n \hat{R}(\lambda)_{ij}} = 
\frac{\sum_{j=1}^n R_{jk}- \gamma_k(\mathbf{\tilde{c}})}{\sum_{j=1}^n \sum_{i=1}^n R_{ij}} = \frac{\sum_{j=1}^n R_{jk}}{\sum_{j=1}^n \sum_{i=1}^n R_{ij}} - \frac{\gamma_k(\mathbf{\tilde{c}})}{\sum_{j=1}^n \sum_{i=1}^n R_{ij}},
\end{align*}

where $\gamma_k(\mathbf{\tilde{c}}) = \det(\mathbf{\tilde{c}}, \mathbf{c}_2-\mathbf{c}_1,...,\mathbf{c}_{k-1}-\mathbf{c}_1,\mathbf{1},\mathbf{c}_{k+1}-\mathbf{c}_1,...,\mathbf{c}_n-\mathbf{c}_1)$. Note that if the matrix $R$ is row-sum-constant, that is, $\mathbf{\tilde{c}}=\nu \mathbf{1}$ for some real $\nu$, then $\gamma_k(\mathbf{\tilde{c}})=0,\forall k$. Hence, in view of (\ref{xk}) and (\ref{xk2}), $\tilde{x}_k = \tilde{x}_k(\lambda)$, which is a contradiction.

Let us now check that $\tilde{\mathbf{x}}$ is also an ESS for $\Gamma_\lambda$. We require that $\mathbf{y} \hat{R}(\lambda) \tilde{\mathbf{x}} < \tilde{\mathbf{x}} \hat{R}(\lambda) \tilde{\mathbf{x}},\forall \mathbf{y}$. Note that

$$\mathbf{y} \hat{R}(\lambda) \tilde{\mathbf{x}} = \mathbf{y} R \tilde{\mathbf{x}} + \frac{1-\lambda}{\lambda n} \mathbf{y} RJ\tilde{\mathbf{x}},$$

$$\tilde{\mathbf{x}} \hat{R}(\lambda) \tilde{\mathbf{x}} = \tilde{\mathbf{x}} R \tilde{\mathbf{x}} + \frac{1-\lambda}{\lambda n} \tilde{\mathbf{x}} RJ\tilde{\mathbf{x}}.$$

As $\tilde{\mathbf{x}}$ is an interior ESS for $R$, we have $\mathbf{y} R \tilde{\mathbf{x}} < \tilde{\mathbf{x}} R \tilde{\mathbf{x}},\forall \mathbf{y}$. Furthermore, $R\tilde{\mathbf{x}} = (\tilde{\mathbf{x}} R \tilde{\mathbf{x}}) \mathbf{1}$ and $RJ\tilde{\mathbf{x}} = R \mathbf{1}=\nu\mathbf{1}$, where $\nu$ is the sum of any row of $R$. Hence

$$ \mathbf{y} RJ\tilde{\mathbf{x}} = \mathbf{y} R \mathbf{1} = \nu = \tilde{\mathbf{x}}RJ\tilde{\mathbf{x}}.$$

Combining the above we obtain

$$\mathbf{y}R\mathbf{\tilde{x}}+\frac{1-\lambda}{\lambda n} \mathbf{y}RJ\mathbf{\tilde{x}} < \mathbf{\tilde{x}}R\mathbf{\tilde{x}}+\frac{1-\lambda}{\lambda n} \mathbf{\tilde{x}}RJ\mathbf{\tilde{x}}.$$

In the same manner we are obtaining the second condition on ESS.

\hfill $\square$

The above result postulates that if initial incompetence is uniform, then the effect of mistakes is neglected in a row-sum constant game, which induces no overall fitness advantage to any strategy. In other words, if in a row-sum constant game everyone is making the same mistakes with the same probabilities, then population dynamics is invariant under these mistakes. We can extend this result for the general form of the fitness matrix as follows:

\begin{theorem}
\label{interiorESS}
Let $\tilde{\mathbf{x}}$ be an interior ESS of $\Gamma_1$. If the starting level of incompetence, $S$, is a uniform matrix, that is, $s_{ij} = \frac{1}{n},\forall i,j=1,\ldots,n$, then 
\begin{equation}
\mathbf{\tilde{x}}(\mu)=\frac{1}{1-\mu} \left( \mathbf{\tilde{x}} - \frac{\mu}{n} \mathbf{1} \right),
\end{equation} 
where $\mu=1-\lambda$, is an interior ESS for the incompetent game $\hat{R}(\lambda)$ and $\mu$ is sufficiently close to $0$. 
\end{theorem}

{\bf Proof:}
For the uniform game we will consider the simplified version with $\hat{R}(\lambda)=R+\frac{1-\lambda}{\lambda n} RJ$. For $\mu=1-\lambda$ being sufficiently close to $0$ we obtain a game which possesses an ESS \cite{Bomze1983}. Then, this ESS can be found as
$$ \mathbf{\tilde{x}}(\lambda)=\phi(\lambda)\hat{R}(\lambda)^{-1}\mathbf{1},$$
where $\phi(\lambda)$ is as in (\ref{IncompMeanFitness}). Hence, we need to analyze the inverse of the incompetent fitness matrix. It can be done as follows:
$$\tilde{R}(\lambda)^{-1}=\hat{R}(1-\mu)^{-1} = \left[ R \left( I - \frac{\mu}{(\mu-1) n}J \right) \right]^{-1}.$$
Let $\frac{\mu}{(\mu-1) n}J = W$, then we can apply Neumann series to obtain the inverse of $(I-W)$ as follows
$$\hat{R}(\mu)^{-1} = \left( I + W + W^2 + W^3 + ... \right) R^{-1}.$$
Then, the ESS for the incompetent game can be defined as
$$ \mathbf{\tilde{x}}(\mu)=\phi(1-\mu) \left( I + \frac{\mu}{(\mu-1)n}J + \left( \frac{\mu}{(\mu-1)n}J \right)^2 + \left( \frac{\mu}{(\mu-1)n}J  \right)^3 + ...\right) R^{-1}\mathbf{1},$$
which in turn can be simplified due to the fact that $\mathbf{\tilde{x}} = \phi(1) R^{-1} \mathbf{1}$, $J^k = n^{k-1}J$ and that we obtain a geometric series in the parenthesis, hence,

$$\mathbf{\tilde{x}}(\mu) = \frac{\phi(1-\mu)}{\phi(1)} \left( I + \frac{\mu}{(\mu-1)n} \times \frac{1}{1-\frac{\mu}{\mu-1}} J \right) \mathbf{\tilde{x}} = \frac{\phi(1-\mu)}{\phi(1)} \left( I - \frac{\mu}{n} J \right) \mathbf{\tilde{x}}.$$
Notice that $\phi(1) = \frac{det(R)}{\sum_i \sum_j R_{ij}}$, $\phi(1-\mu) = \frac{det(\hat{R}(1-\mu))}{\sum_i \sum_j R_{ij}}$ and hence by Lemma \ref{determinant} the proof is complete.

\hfill $\square$

Note that, the point $\mathbf{\tilde{x}}(\mu)=\frac{1}{1-\mu} \left( \mathbf{\tilde{x}} - \frac{\mu}{n} \mathbf{1} \right)$ remains a fixed point of the replicator dynamics as long as it lies in the interior of the simplex, that is, as long as $\tilde{x}_k > \frac{\mu}{n}$. Hence, this point will remain in the interior for all $\mu\in[0,1]$ only for the case when $\tilde{x}_k=\frac{1}{n},\forall n$. 

Let us demonstrate these results on the example of the well-known Rock-Scissors-Paper type game. We consider this game with the matrix R given as follows:

$$\begin{blockarray}{cccc}
& Rock & Scissors & Paper\\
\begin{block}{c(ccc)}
Rock&0&2&-1\\
\\
Scissors&-1&0&3\\
\\
Paper&2&-2&0\\
\end{block}
\end{blockarray}.$$

We assume that the starting level of incompetence is uniform, that is, $S=\frac{1}{3}J$. We see that the interior equilibrium of the original (fully competent) game (see Figure 5, panel (e)) is pushed to the population adopting only Scissors strategy (see Figure 5, panel (a)) in the case with the uniform starting level of incompetence which is determined by the payoff matrix structure. Then, as the incompetence parameter, $\lambda$, is growing, that is, competence is increasing, we see this equilibrium moving back towards the interior of the simplex (see Figure 5, panels (b)-(d)).

\begin{figure}[h!]
\begin{center}
\includegraphics[scale=0.45]{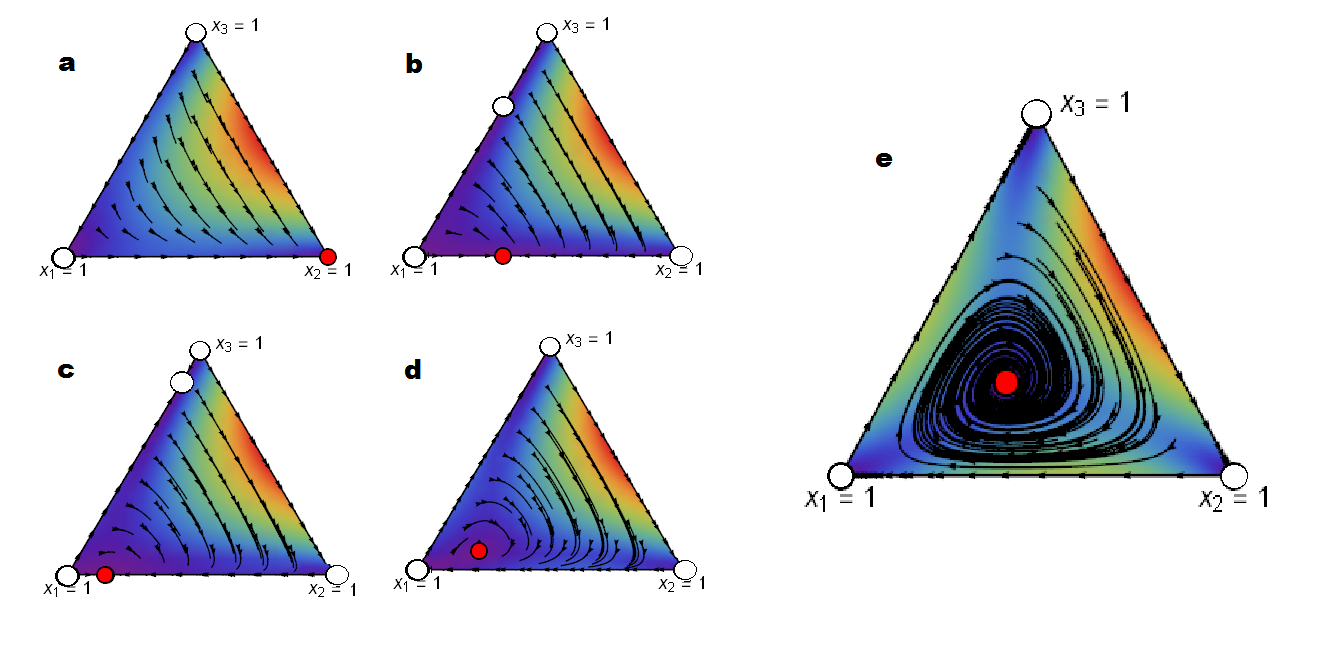}
\end{center}
\caption{Game flow for the Rock-Scissors-Paper game with (a) $\lambda=0.1$ (b) $\lambda = 0.2$ (c) $\lambda = 0.226$ (d) $\lambda = 0.3$ (e) $\lambda = 1$.}
\end{figure}

Theorem \ref{interiorESS} implies that we adjust the ESS of the original game depending on the level of incompetence. Furthermore, we may say that no learning advantages make the population resistant to environmental changes. On the other hand, different adaptability of strategies may be beneficial for species as they can randomize their reactions and re-adapt to new conditions. For instance, some strategies may become preferable under specific conditions. Furthermore, incompetence may be used to estimate perturbations in environmental conditions that violate evolutionary stability of population behavioral choice.


\section{Conclusions}
\label{summary}

In this paper, we explored the influence of environmental fluctuations on the social behavior of species. Depending on the form of fluctuations and their frequency, population may react in a different manner. However, in Section 4, we observed that if changes are providing advantages to some strategies, then asymmetry in reactions may arise. 

In addition, in Section 3, we examined how individual adaptation process of species influences the population dynamics. Furthermore, we determined the critical time before which the population is in the evolutionarily weak phase and can be invaded, as its competence level has not yet reached the maximal critical value. This becomes even more important when invasions happen in between seasons in the case with periodic fluctuations when species are adapting to new conditions. 

However, this weak phase effect can be neglected when environmental changes imply no asymmetry in strategic learning, that is, in the case with uniform starting incompetence (see Lemma 5). However, it is natural to assume that in most cases we would expect that specific environmental conditions require different strategies. This in turn makes even evolutionary stable population vulnerable when facing invasions during the adaptation. We can even say that in some cases learning advantages make a population resilient to environmental changes. 

We explored the population dynamics under incompetence when a population reaches the evolutionary stable state and we estimated a time interval required for the adaptation process. However, depending on the environmental shifts and the populations under consideration this time interval can be long. Thus, properties of the system during the weak phase could be important. Hence, it will be interesting to explore the behavior of the population dynamics under incompetence and their resilience to invasions. Furthermore, this idea can, perhaps, be fruitfully extended to the games where two or more populations interact.

\bibliographystyle{plain}
\bibliography{MyRefs}

\end{document}